\documentclass[12pt]{iopart}

\usepackage{graphicx}
\expandafter\let\csname equation*\endcsname\relax
\expandafter\let\csname endequation*\endcsname\relax
\usepackage{amsmath}
\usepackage{iopams}
\usepackage{amssymb}
\usepackage{bm}

\newcommand{\dd}{\;\mathrm{d}} 

\newcommand{\eq}[1]{Eq.~\ref{eq:#1}}

\newcommand{\nb}[1]{}
\newcommand{\nba}[1]{}
\newcommand{\nbold}[1]{}

\begin{document}

\title[Ad-hoc data correction approach for reliable atomic pair distribution functions]{Towards a robust \emph{ad-hoc} data correction approach that yields reliable atomic pair distribution functions from powder diffraction data}

%
\author{Simon~J.~L. Billinge}
\address{Department of Applied Physics and Applied Mathematics, Columbia
University, New York, NY, 10027, USA}
\address{Condensed Matter Physics and Materials Science Department,
Brookhaven National Laboratory Upton, NY 11973, USA}
\ead{sb2896@columbia.edu}
\author{Christopher L. Farrow}
\address{Department of Applied Physics and Applied Mathematics, Columbia
University New York, NY, 10027, USA}


\begin{abstract}
We examine the equations to obtain atomic pair distribution functions (PDFs) from x-ray, neutron and electron powder diffraction data with a view to obtaining reliable and accurate PDFs from the raw data using a largely \emph{ad hoc} correction process.  We find that this should be possible under certain circumstances that hold, to a reasonably good approximation, in many modern experiments.  We describe a variational approach that could be applied to find data correction parameters that is highly automatable and should require little in the way of user inputs yet results in quantitatively reliable PDFs, modulo unknown scale factors that are often not of scientific interest when profile fitting models are applied to the data with scale-factor as a parameter. We have worked on a particular implementation of these ideas and demonstrate that it yields PDFs that are of comparable quality to those obtained with existing x-ray data reduction program PDFgetX2. This opens the door to rapid and highly automated processing of raw data to obtain PDFs.

\end{abstract}


\maketitle

\section{Introduction}
\label{sec:introduction}

Total scattering analysis and the closely related atomic pair distribution function (PDF) method, are growing in popularity in the area of nanostructure determination~\cite{billi;cc04,billi;jssc08,young;jmc11}, where the PDF is the Fourier transform of the total scattering structure function. Total scattering data and the PDF can be obtained from x-ray, neutron~\cite{warre;b;xd90,egami;b;utbp03,egami;b;utbp12} and electron~\cite{cocka;aca88,abeyk;zk12} data from isotropically scattering samples such as crystalline powders, nanoparticles, amorphous materials and liquids.  It contains information about structure at the nanoscale since it utilizes both Bragg and diffuse scattering intensities which contribute information about the average and local structures, respectively.  With the advent of high power x-ray and neutron sources with optimized PDF instruments~\cite{chupa;jac03,proff;apa01i}, the emerging realization that for nanoparticles quantitatively reliable PDFs can be obtained from electron diffraction data~\cite{abeyk;zk12}, and the maturing of sophisticated computer based modeling programs~\cite{proff;jac99,farro;jpcm07,proff;jac97,tucke;jpcm07,soper;molp01,cerve;jac10}, the use of the PDF is expected to have a significant impact in the area of nanostructure characterization in the coming years.

In order to obtain the total scattering structure function and the PDF from the data, significant corrections have to be made to the raw data, as well as proper normalization, before Fourier transforming to obtain the PDF, $G(r)$, as described in detail in Chapter 5 of~\cite{egami;b;utbp03,egami;b;utbp12}.  Computer programs exist for doing this~\cite{qiu;jac04i,petko;jac89,jeong;jac01,peter;jac00}\nba{add gudrun reference}, but it remains a tedious and problematic process and a barrier to broader adoption of the method.  Here we investigate whether quantitatively reliable PDFs can be obtained from powder diffraction data using purely \emph{ad hoc} corrections.  ``Quick and dirty'' PDFs have been obtained in this way for some time~\cite{egami;b;utbp03}.  We explore this in more detail and show that, if certain experimental conditions hold, not only ``quick and dirty" but \emph{accurate} PDFs may be obtained using completely \emph{ad hoc} correction methods and we propose a variational approach that should allow quantitatively correct reduced structure functions and PDFs to be obtained this way, modulo a global scale factor on the intensities and the atomic displacement parameters.  We have implemented this approach in a program, the details of which will be described elsewhere.  However, we reproduce a figure here that direct demonstrates the promise of this approach by yielding x-ray PDFs of comparable quality to those obtained from the widely used PDFgetX2 program~\cite{qiu;jac04i}. This has the potential to greatly simplify total scattering and PDF studies, for example, facilitating real-time data processing during data collection.

\section{data reduction}
\label{sec:reduction}

We will first discuss how the total scattering
structure function is typically obtained in x-ray and neutron diffraction. This
process has an established theoretical foundation~\cite{egami;b;utbp03}.

\subsection{X-ray and neutron diffraction data reduction}

The reduced total scattering structure function, $F(Q)$, is defined in terms of
the total scattering structure function, $S(Q)$, as
\begin{equation}
F(Q) = Q(S(Q)-1).
\label{eq:fq}\end{equation}
The
structure function contains the discrete coherent singly scattered information
available in the raw diffraction intensity data.  It is defined according
to~\cite{warre;b;xd90}
\begin{equation}
\label{eq:sofq}
S(Q) = \frac{I_c(Q)}{N \langle f\rangle^2} - \frac{ \langle (f - \langle f
\rangle)^2 \rangle}{\langle f \rangle^2},
\end{equation}
which gives~\cite{farro;aca09}
\begin{equation}
\begin{split}
S(Q) - 1 =& \frac{I_c(Q) - N \langle f^2\rangle}{N\langle f\rangle^2} \\
=& \frac{I_d(Q)}{N\langle f\rangle^2},
\end{split}
\label{eq:sofqm1}
\end{equation}
where $f$ is the $Q$-dependent x-ray or electron scattering factor or $Q$-independent
neutron scattering length, as appropriate, and $\langle \ldots \rangle$
represents an average over all atoms in the sample.  In this equation, $I_c(Q)$
is the coherent single-scattered intensity per atom and $I_d(Q)$ is the
discrete coherent scattering intensity, which excludes the self-scattering,
$N\langle f^2\rangle$~\cite{farro;aca09}.  The coherent scattering intensity is
obtained from the measured intensity by removing parasitic scattering (e.g.,
from sample environments), incoherent and multiple scattering contributions,
and correcting for experimental effects such as absorption, detector
efficiencies, detector dead-time and so on~\cite{egami;b;utbp03}. The resulting
corrected measured intensity is normalized by the incident flux to obtain
$I_c(Q)$. The self-scattering, $N\langle f^2\rangle$, and normalization,
$N\langle f\rangle^2$, terms are calculated from the known composition of the
sample using tabulated values of $f$.

As evident in \eq{sofq}, to obtain $S(Q) - 1$ from $I_c(Q)$ we subtract the
self-scattering, $N\langle f^2\rangle$, which has no atom-pair correlation
information, and divide by $N\langle f\rangle^2$.  As a result, $S(Q) - 1$
oscillates around zero, and asymptotically approaches it at high~$Q$ as the
coherence of the scattering is lost.  If the experimental effects are removed
correctly, the resulting $F(Q)$ and $G(r)$ are directly related to, and can be
calculated from, structural models~\cite{farro;aca09}. The corrections are well
controlled in most cases and refinements of structural models result in reduced
$\chi^2$ values that approach unity in the best cases.  Some uncertainty in the
corrections can be tolerated.  This is due to a somewhat fortuitous circumstance
that they are mostly long-wavelength in nature, such as the Compton scattering correction
in the case of x-rays, whereas the signal from the structure is much higher frequency in $Q$.
If these long wavelength contributions are not correctly
removed they result in correspondingly long-wavelength aberrations to $S(Q)$ that
appear in $G(r)$ as peaks in the very low-$r$ region below any physically meaningful PDF peaks~\cite{peter;jac00}.

There are various programs for obtaining PDF from raw data,
such as PDFgetX2~\cite{qiu;jac04i}, RAD~\cite{petko;jac89} and GudrunX~\footnote{available from the ISIS disordered
materials group website, http://http://www.isis.stfc.ac.uk/instruments/sandals/data-analysis/gudrun8864.html}
for x-rays and PDFgetN~\cite{peter;jac00} and Gudrun$^1$
for time of flight neutrons.  These programs provide excellent results but require numerous data inputs and user interactions and are difficult to learn, with multi-day workshops sometimes being dedicated to learning their use.  If a much simpler data correction protocol could be found that resulted in PDFs of comparable quality, but which could be automated, it would potentially greatly expand and assist the PDF community.  Here we explore whether a protocol can be found using completely \emph{ad hoc} corrections that can result in quantitatively reliable PDFs.  To be explicit, we seek the actual reduced structure function $F(Q)$ from a sample, given the measured powder diffraction, $I_m(Q)$.  In a conventional data reduction, we begin by finding the coherent scattered intensity, $I_c(Q)$ from $I_m(Q)$ by making corrections for things such as detector deadtime, polarization, multiple scattering, backgrounds, and so on.  The reduced structure function is then determined from Eqs.~\ref{eq:fq}, \ref{eq:sofq}, and \ref{eq:sofqm1}.

Apart from the detector dead-time correction, all the corrections are either simply additive or multiplicative.  If we assume that any detector dead-time is negligible or has been corrected before getting $I_m$, we can write
\begin{equation}
I_c = a(Q)I_m(Q) + b(Q),
\end{equation}
Where $a$ and $b$ are the generalized (and unknown) $Q$-dependent multiplicative and additive, respectively, correction functions.  It is these additive and multiplicative corrections that are explicitly calculated from theory~\cite{egami;b;utbp03} and applied in the PDF data reduction programs mentioned above based on detailed user inputs about the experimental conditions.

Careful inspection of Eqs.~\ref{eq:fq}, \ref{eq:sofq}, and \ref{eq:sofqm1} shows that we can also write an expression for $F(Q)$ itself in the same form.
\begin{equation}
F(Q) = \alpha (Q) I_m(Q) + \beta (Q)
\end{equation}
without loss of generality.

Writing the equations this way is of no particular advantage because we don't know the form, or the $Q$-dependence, of $\alpha$ and $\beta$.  However, we do have considerable information about the nature and asymptotic behavior of $F(Q)$ and we do have some information about the nature of the physical corrections that combine to make $\alpha$ and $\beta$.  Here we show how, in principle, this can be used to determine the properly corrected $F(Q)$ with minimal input information.

Careful consideration of the nature of  the structural and non structural components to the measured signal suggests that there is a good separation between the frequency of most corrections and the frequency of the structural information in the PDF.  The lowest frequency Fourier component in $F(Q)$ coming from a real structural signal is $\sim 2\pi/r_{nn}$ where $r_{nn}$ is the length of the shortest inter-atomic bond-length. This means that all \emph{additive} frequency components in the signal that have lower frequency than this are certainly coming from non-structural contributions to the signal.  On the other hand, as we discussed above, the additive contributions to the signal coming from extrinsic sources are predominantly much longer wavelength and more slowly varying than this.

If we assume for the moment that the multiplicative corrections have all been
correctly applied to $I_m$ (i.e., set $\alpha (Q)$ to unity) we could fit a
smooth curve that has only frequency components higher than $2\pi/r_{nn}$ through the data and subtract it.  This will result in a function that has the correct asymptotic behavior as $F(Q)$, oscillating around zero, and actually is $mF(Q)$ if there are no experimental aberrations with frequency components  higher than $2\pi/r_{nn}$, where $m$ is an unknown constant that affects the scale of the resulting $F(Q)$ but not its shape.  A similar approach has been used for many years as a \emph{post-facto} correction to clean-up unwanted oscillations in the low-$r$ region of the PDF. In this approach the low-$r$ ripples are back-Fourier transformed to $Q$-space and then this signal is subtracted from the $F(Q)$ before again Fourier transforming the corrected $F(Q)$ resulting in a cosmetically improved PDF.  However, here we argue that a completely \emph{ad hoc} correction can give PDFs of comparable quality to those obtained by traditional approaches but with much less user and computational effort, modulo an unknown scale factor.  This last fact means that other information must be used to obtain the correct absolute scale for the data. However, in many cases this information available from other sources and structure refinement programs such as PDFgui refine scale factor as a variable and it is not fixed in any case.

This low-frequency requirement holds in practice very well for all the additive corrections except for those that actually contain structural information themselves, such as scattering from the sample container.  However, if sample container scattering is significant it can be measured and subtracted fairly straightforwardly.

\begin{figure}
\includegraphics[width=\columnwidth]{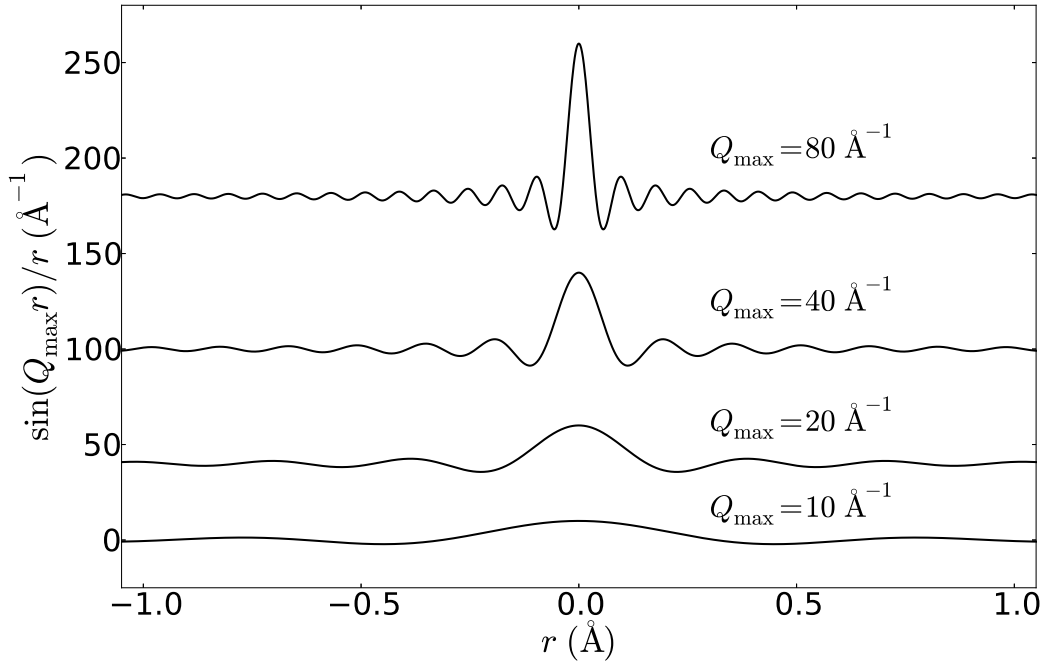}
\caption{\label{fig:sinc}
The modified sinc function as it appears in the PDF equation, \eq{fronepeak},
for various values of $Q_{\mathrm{max}}$.
}
\end{figure}

We now consider the effect on the PDF of the multiplicative term, $\alpha(Q)$.
For reference, consider the ideal correlation function, $G_{ij}(r_{ij})$ from a single atom pair
$(i, j)$ situated a distance $r_{ij}$ apart. Using Debye's
equation~\cite{debye;ap15} for the coherent scattering amplitude,
\begin{equation}
\label{eq:Idebye}
I_c(Q) = \sum_i \sum_j f_i f_j^* \frac{ \sin (Qr_{ij})}{Qr_{ij}},
\end{equation}
 we get $F_{ij}(Q)$ corresponding to the single peak PDF as
\begin{equation}
\label{eq:smsonepeak}
F_{ij}(Q) \propto \frac{\sin(Q r_{ij})}{r_{ij}}.
\end{equation}
This gives for the PDF~\cite{farro;aca09},
\begin{equation}
\label{eq:fronepeak}
\begin{split}
G_{ij}(r) \propto &\int_{Q_{min}}^{Q_{max}} \frac{\sin(Q r_{ij})}{r_{ij}} \sin(Qr) \dd Q \\
    = & \frac{\sin((r - r_{ij})Q_{max})}{r-r_{ij}} -
         \frac{\sin((r + r_{ij})Q_{max})}{r+r_{ij}} \\
    - & \frac{\sin((r - r_{ij})Q_{min})}{r-r_{ij}} +
         \frac{\sin((r + r_{ij})Q_{min})}{r+r_{ij}},
\end{split}
\end{equation}
which is the sum of two signals, one with maximum at $r_{ij}$, and the other at
$-r_{ij}$.  We ignore the contribution from $Q_{min}$, which oscillates much
slower than the contributions from $Q_{max}$.
In general we only compute PDF on the  positive axis and
the contribution on the positive axis to the peak centered at $-r_{ij}$ is ignored with little
loss in accuracy as its contribution on the positive axis is
small~\cite{thorp;b;fstpbas02}. As we expect, the PDF is a peak at the position
$r_{ij}$ but with the characteristics of a Sinc function due to the finite
Fourier transform.  The cental peak of the Sinc function has a FWHM that is inversely proportional to the width of the window in $Q$-space \nba{clf:I can't find a fwhm
number for sinc. It's a transcenental function, so it probably has a nice
elegant solution that is way beyond my skills.} with intensity tails that die off as $1/r$ away from the peak on the low- and high-$r$ sides modulated by an oscillation with a wavelength of $1/Q_{max}$.  It is shown in Fig.~\ref{fig:sinc} for various values of $Q_{max}$.
For large $Q_{max}$ these signals approach Dirac-delta functions centered at $\pm r_{ij}$. Without taking into consideration peak broadening
due to thermal fluctuations, the finite width of these peaks is solely due to
the finite $Q_{max}$.

Next we consider the effect of multiplicative distortions $\alpha^\prime(Q)$ on this signal.  In other words,
we assume that the multiplicative corrections have not been done correctly and we Fourier transform
\begin{equation}
F^\prime (Q) = \alpha^\prime (Q) F(Q)
\end{equation}
instead of $F(Q)$ itself.

In the best case, $\alpha^\prime(Q)$ is constant
and it scales the peaks of the correlation function uniformly, which does not
distort the structural information.  Models fit to data that is distorted only by a constant scale factor gave equivalent structural results provided a constant scale-factor could be refined in the model~\cite{peter;jac00}.

Now let us consider the effects on the PDF, $G(r)$, of a $Q$-dependent $\alpha^\prime(Q)$.
To do this we assume that $\alpha^\prime(Q)$ has a convergent Fourier
series expansion over the interval $[0, Q'_{max}]$, and that $Q'_{max} \ge
Q_{max}$. This means that the longest wavelength component of $\alpha(Q)$ may be
greater than the extent of the measured signal.  We express the Fourier
expansion of $\alpha^\prime(Q)$ as
\begin{equation}
\label{eq:falpha}
\begin{split}
\alpha^\prime(Q) &= \frac{a_0}{2} + \sum_{n=1}^{\infty}
    \left(
    a_n \cos\left(\frac{2 \pi n}{Q'_{max}} Q\right)
    +
    b_n \sin\left(\frac{2 \pi n}{Q'_{max}} Q\right)
    \right)\\
    &= \frac{a_0}{2} + \sum_{n=1}^{\infty}
    \left(
    a_n \cos\left(r_n^\prime Q\right)
    +
    b_n \sin\left(r^\prime_n Q\right)
    \right),\\
\end{split}
\end{equation}
which serves to define
$r_n'$. Long wavelength Fourier
components in $\alpha^\prime(Q)$ correspond to small values of $r_n'$.  We only need to consider
the cosine components of $\alpha^\prime(Q)$, because the sine components do not
contribute to $G(r)$ due to the sine Fourier transform.  Thus, for a given $n$, we
have for our $F^\prime(Q)$ of a single peak PDF
\begin{equation}
\begin{split}
F^\prime(Q) &\propto a_n \cos(r_n' Q) \frac{\sin(Q r_{ij})}{r_{ij}} \\
      &= \frac{a_n}{2} \left[ \frac{\sin(Q(r_{ij}+r_n'))}{r_{ij}}
                   +\frac{\sin(Q(r_{ij}-r_n'))}{r_{ij}}\right].\\
\end{split}
\end{equation}
Here, we have used trigonometric identities to go from the first line to the
second line. Putting this into the form of \eq{smsonepeak},
\begin{equation}
\begin{split}
\label{eq:smsonepeakmod}
F(Q) &\propto \frac{a_n}{2} \left(1 + \frac{r_n'}{r_{ij}}\right)
             \frac{\sin(Q(r_{ij}+r_n'))}{r_{ij} + r_n'}\\
      &+ \frac{a_n}{2}\left(1 - \frac{r_n'}{r_{ij}}\right)
             \frac{\sin(Q(r_{ij}-r_n'))}{r_{ij} - r_n'}.
\end{split}
\end{equation}

Comparing \eq{smsonepeakmod} to \eq{smsonepeak}, we see that instead of a
single peak at $r_{ij}$ in $G(r)$, \eq{smsonepeakmod} produces two peaks of almost equal
intensity, one
at $r_{ij}+r_n'$ and one at $r_{ij}-r_n'$. In actuality, the precise amplitudes of these
two peaks are not the same; the peak at $r_{ij}+r_n'$ is larger than the one at
$r_{ij}-r_n'$, and the amplitude difference is $2r_n'/r_{ij}$.

As we discussed earlier, we expect most aberrations coming from imperfect multiplicative
corrections to be long-wavelength, for example, extinction and absorption corrections.
These only have Fourier components with small $r'_n$, in the limit $r_n' \ll r_{ij}$.
In this case, the two distinct Sinc peaks would appear as a single unresolved but broadened peak close to the position of the undistorted PDF peak at $r_{ij}$.
As $r_n'/r_{ij}$ gets larger, the peak would further broaden, shift slightly and become asymmetric due to the
amplitude difference of the signals. The position of the
maximum of an asymmetric peak would be larger than $r_{ij}$ due to this
asymmetry, and asymmetry would be more pronounced for peaks at lower $r$.  The
combined influence from multiple Fourier components would smear out any peak
splitting but accentuate the asymmetry of the peak.
Thus, the effect of all imperfectly
corrected multiplicative aberrations to $F(Q)$ is to broaden, and skew peaks in the PDF.
We are familiar with this peak broadening effect as the same effect as given by the convolution theorem of Fourier transforms.  The peak in $r$-space is being broadened by the convolution of the Fourier transform of $\alpha^\prime (Q)$ with the pristine PDF peak.

Because all long-wavelength aberrations in the data result in a broadening of the peaks, this suggests that an {\it ad hoc} variational approach could be found to search for the Fourier coefficients
of some unknown $\alpha~\prime (Q)$ where these are adjusted in such a way as to make the resulting PDF peaks as sharp and as symmetric as they can be. This could be automated in a regression scheme.  A challenge here is that there is also peak broadening in the data that has real physical significance: the thermal motions and static disorder.  Indeed, this broadening is produced by a low-frequency multiplicative factor applied to the intensity in $Q$-space, the
Debye-Waller factor~\cite{warre;b;xd90}.  Thus, unlike the case of the additive corrections, there is not a clean separation in frequency of the physical signal and the experimental aberrations that we can exploit here. It is possible that a scheme could be found to separate the contributions by applying some additional knowledge about the behavior of the different functions.  For example, the Debye-Waller factor affects different PDF peaks differently depending on the atoms contributing to the peak whereas the data corrections do not have this chemical specificity.  Thus, the relative atomic displacement factors could be recovered modulo an uncertain overall scale that may be obtained from other measurements.
\nba{clf:I hate leaving this at a cliffhanger. I would love to put a working algorithm
in this paper, but who knows how long that would take to develop. Since we know
that the DW factor convolves the Sinc PDF peaks with a Gaussian, the unknown
$\alpha(Q)$ can be determined via deconvolution. So, on the down side, it's a
deconvolution. On the up side, there's a ton of research into deconvolution.
(BTW, did you see the photoshop deblurring demo. Proof that deconvolution can
be done.)
}

To test out these ideas we have created an implementation of the procedure.  The program is described in detail elsewhere~\cite{juhas;unpub12}.  It models $\beta^\prime (Q)$ as an polynomial of no more than 8 or 9 orders, where this number is chosen to ensure limit the highest frequency possible in $\beta^\prime (Q)$~\cite{juhas;unpub12}.  At this time $\alpha^\prime (Q)$ is simply set to unity.  This implementation has been tested on real data obtained in high energy rapid acquisition PDF mode on fine powders and nanomaterials.  Under these conditions the absorption and extinction effects are expected to be small.  The data were reduced from raw 1D intensities to PDFs using the new procedure, and these PDFs were compared to the PDFs obtained from the same data using the PDFgetX2 program~\cite{qiu;jac04i}.  The comparisons are very good.  The result for a representative sample of Ni is shown in Figure~\ref{fig;comp}.
\begin{figure}
\includegraphics[width=\columnwidth]{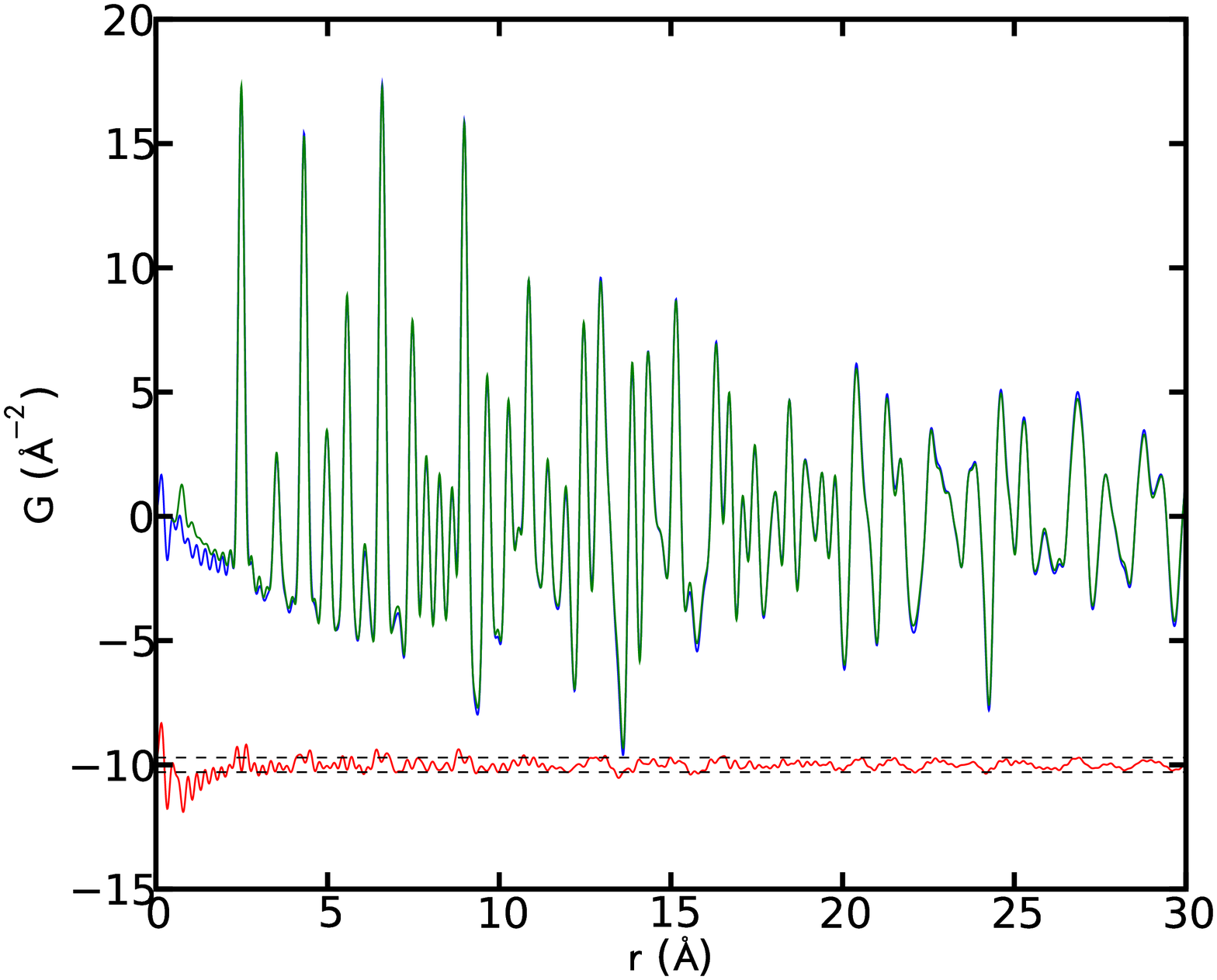}
\caption{\label{fig:comp}
Comparison of PDFs obtained using the new procedure and using PDFgetX2 which implements all the corrections explicitly.
The data are from a high energy x-ray measurement of nickel powder.  The PDF obtained with the \emph{ad hoc} procedure is plotted in green and the one obtained using PDFgetX2 is shown in blue.  The difference curve is offset below.  The horizontal dashed lines are guides to the eye. See~\cite{juhas;unpub12} for more details.
}
\end{figure}
For more details and more comparisons we point the reader to the publication on the program, PDFgetX3~\cite{juhas;unpub12}.  Nonetheless, this comparison shows that this \emph{ad hoc} approach to data corrections works rather well in practice and may be used to obtain quantitatively reliable PDFs, at least under favorable experimental conditions.

\section{Acknowledgments}

We would like to thank Pavol Juh\'{a}s, Timur Dykhne and Emil Bo\v{z}in for allowing us to reproduce Figure~2, as
well as for useful discussions.
This work was supported by the US National Science foundation through Grant
DMR-0703940.

\bibliography{abb-billinge-group,everyone,billinge-group,120715notes}
\bibliographystyle{unsrt}

\end{document}